\documentstyle[12pt]{article}          % XXX preprint style
\newcommand{\be}{\begin{equation}}
\newcommand{\ee}{\end{equation}}
\newcommand{\ba}{\begin{eqnarray}}
\newcommand{\ea}{\end{eqnarray}}

\begin{document}

\baselineskip 15pt
\parindent=1cm
\parskip 3mm

\title{The dependence of strange hadron multiplicities \\
 on the speed
of hadronization}

\author{
J. Zim\'anyi$^{\star}$,
T.S. Bir\'o and P. L\'evai \\ \\
 KFKI Research Institute for Particle and Nuclear Physics, \\
POB. 49, Budapest, 1525, Hungary \\
}

\footnotetext{$^\star$ The talk was presented by J. Zim\'anyi
on the Strangeness'97 Conference, Santorini, April 14-18 1997.
E-mail: jzimanyi@sunserv.kfki.hu}

\date{June 12 1997}

\maketitle

\begin{abstract}
Hadron multiplicities are calculated in the ALCOR
model \cite{ALCOR, ALCORS95, ALCORS96} for 
the Pb+Pb collisions at CERN SPS energy. 
Considering the newest experimental results, we display our prediction
obtained from the ALCOR model for stable hadrons including strange
baryons and anti-baryons. 
\end{abstract}

\section{Introduction}

With the presentation of this talk 
we would like to raise the question how a 
theory can be disproved.
To reply this question we have to point out that quantitative
predictions of a model must be published before the corresponding
experimental data are known. In this spirit we published our
predictions for the Pb+Pb collision in the proceedings of
Strange Matter '96 \cite{ALCORS96}.
Now it turned out that the simple ALCOR model, which
described excellently the S+S results \cite{ALCORS95,ALCORS96}
fails to account for all the recently measured data in
Pb+Pb collision, presented at this conference \cite{NA49S97,WA97S97}.
Therefore at least one of the assumptions underlying the
original ALCOR model has to be modified.
In particular searching for the cause of our disagreement
on the $\overline{\Omega}^+/\Omega^-$ ratio
we found that not all hadronization processes are as
sudden as it would verify the coalescence approximation
of the underlying evolution.
\medskip

%\vspace{0.3cm}
Since  a long time the strange particles are considered to be
the important messengers of the happenings in the heavy ion
reactions. This has two aspects:
i) the amount of created $S \overline S $ pairs and
ii) the ratio of the multiplicities of different strange
hadrons.
While the first question rather belongs to the area of the 
collision models,
rehadronization models search the answers to the second
problem.
\medskip
 
%\vspace{0.3cm} 
In the  early years it was assumed that by whatever mechanism,
the colliding nuclei will  stop on each other to produce a
 quark-gluon plasma. It was assumed to exist in thermal and
chemical equilibrium. At the next level of theoretical
sophistication  it was assumed to
{ hadronize}  in a very slow { quasi-equilibrium
process}. When compared to the pure hadronic scenario, 
this leads to a strongly
different multiplicity distribution. Then we were happy that we
had a { nice signature} for the quark-gluon plasma.
\medskip
%\newpage

%\vspace{0.3cm} 
In the last six years, however, different nucleon-nucleon
collision models were introduced, which claimed that they can
describe the ultrarelativistic heavy ion reactions by { purely
hadron-hadron collisions}. Unfortunately, the attention was too
much focused on these Monte Carlo  hadron collision programs.
There were only few attempts to discuss the hadronization
using the concepts of { quasi-stationary} thermodynamics.
We started, on the other hand, a { completely out of equilibrium
hadronization approach}. The first model was constructed 
in 1994 \cite{ALCOR}, and I was happy to hear:
\bigskip
 
%\vspace{0.5cm} 
{\em "Finally, there is a real need for more sophisticated
dynamical models of the transition from quark-gluon plasma to
hadronic gas." }
\bigskip

%\vspace{0.5cm} 
This was said in the lecture of B. M\"uller at
the QM'95 conference in Monterey, in January 1995.
\medskip
 
%\vspace{0.5cm} 
In the present  work we discuss principles of the { fast
hadronization} model, { ALCOR}, and compare its predictions for the
multiplicities of strange hadrons
% with the prediction from a { slow} (thermal) hadronization model and
with experimental data.
Eventually we address the possibility, that the very recently 
measured $\overline{\Omega}^+/\Omega^-$ ratio still can
agree with the fast hadronization scenario for mesons
assuming a time-dependent screening length in the quark matter.

\vspace{0.5cm}
 
First we show how { ALCOR} emerges from { requiring} such a
solution of the rate equations for quarks, antiquarks and diquarks,
by which { all free quarks and antiquarks vanish after
hadronization}.

\vspace{0.5cm}
The basic philosophy of ALCOR consists of several assumptions
about the physical conditions prior to hadronization:
 
\begin{itemize}

\item the initial state is a { quark-antiquark plasma}
 (and not a quark-gluon plasma),

\item the constituent particles are dressed having in-medium masses,
 
\item the momentum distributions of valence quarks and
antiquarks correspond to a thermal distribution having
longitudinal and transversal flow, too,

\item there are microscopic descriptions which yield
cross section for the process $ q + \overline{q} \rightarrow hadrons$,
and the produced hadrons are immediately in the hadron phase 
(like in case of a Mott transition),

\item the momentum distribution of produced first generation
hadrons are determined by the valence quark momentum distribution and
by the kinematics of the microprocesses,

\item the question, whether the system will have a long enough
lifetime for allowing further interactions between hadrons
before break up
 to change the momentum distribution is left open.
\end{itemize}

In the original formulation of the ALCOR model we used the
picture of quark redistribution among the final hadrons for
determining internal parameters, the $b$ factors.
Now we generalize this concept by relating these factors
to the time-integral average of products of different
species numbers.

%%%%%%%%%%%%%%%%%%%%%%%%%%%%%%%%%%%%%%%%%%%%%%
%\newpage
 
\noindent
\section{ Hadronization channels}
 
In general we consider many possible hadronization channels. 
Noting the
   reaction channel index by $\nu$,
   the (anti)quark flavor index by $i,j,k$
   the diquark flavor by $(ij)$
   a meson made from quark $i$ and antiquark $j$ by $[ij]$
  and finally a
   baryon made from quark $i$ and diquark $(jk)$ by $[ijk]$, 
we arrive at reactions in different reaction channels $\nu$.
They can be classified by the following types, creating primarily
mesons, (anti)diquarks and (anti)baryons:
\be
 m^{\nu}_i q_i \, + \overline{m}^{\nu}_j \overline{q}_j
 \, \longrightarrow \,
 m^{\nu}_{[ij]} M_{ij} +
 m^{\nu}_{(ij)} D_{ij} +
 m^{\nu}_{(ij)} \overline{D}_{ij}
\ee
with cross section $\sigma^{\nu}_{i+\overline{j} \rightarrow [ij]}$,
\be
 d^{\nu}_i q_i \, + d^{\nu}_j q_j
 \, \longrightarrow \,
 d^{\nu}_{[ij]} M_{ij} +
 d^{\nu}_{(ij)} D_{ij}
\ee
with cross section $\sigma^{\nu}_{i+j \rightarrow (ij)}$,
\be
b^{\nu}_i q_i \, + b^{\nu}_{(jk)} D_{(jk)}
 \, \longrightarrow \,
 b^{\nu}_{[ij]} M_{ij} +
 b^{\nu}_{[ijk]} B_{ijk}
\ee
with cross section $\sigma^{\nu}_{i+(jk) \rightarrow [ijk]}$.
The $m_i^\nu$, $d_i^\nu$, and $b_i^\nu$ are the appropriate
stoichiometric coefficients.
The processes for antihadrons are obtained due to
interchanging quarks and antiquarks, the cross sections 
we consider are the same.
 
\newpage

\noindent
\section{ Effective rates}
 
Using the above stoichiometric coefficients and cross sections
we define the following effective rates for number changing
reactions:

\noindent \ \ \ \ { quark rates}
\ba
Q_{[ij]} &=& \sum_{\nu} m_i^{\nu}
\langle \sigma^{\nu}_{i+\overline{j} \rightarrow [ij]} v \rangle, \qquad
Q_{(ij)} = \sum_{\nu} d_i^{\nu}
\langle \sigma^{\nu}_{i+j \rightarrow (ij)} v \rangle, 
\nonumber \\ 
Q_{[ijk]} &=& \sum_{\nu} b_i^{\nu}
\langle  \sigma^{\nu}_{i+(jk) \rightarrow [ijk]} v \rangle,
\ea
\ \ \ \ { diquark rates}
\ba
D_{[jk]l} &=& \sum_{\nu} m_{(jk)}^{\nu}
\langle \sigma^{\nu}_{j+\overline{l} \rightarrow (jk)
+(\overline{l}\overline{k})} v \rangle, \qquad
D_{(jk)} = \sum_{\nu} d_{(jk)}^{\nu}
\langle \sigma^{\nu}_{j+k \rightarrow (jk)} v \rangle, \nonumber \\
D_{[ijk]} &=& \sum_{\nu} b_{(jk)}^{\nu}
\langle  \sigma^{\nu}_{i+(jk) \rightarrow [ijk]} v \rangle,
\ea
\ \ \ \ { meson rates}
\ba
M_{[ij]} &=& \sum_{\nu} m_{[ij]}^{\nu}
\langle \sigma^{\nu}_{i+\overline{j} \rightarrow [ij]} v \rangle, \qquad
M_{(ij)} = \sum_{\nu} d_{[ij]}^{\nu}
\langle \sigma^{\nu}_{i+j \rightarrow (ij)} v \rangle, \nonumber \\
M_{[ijk]} &=& \sum_{\nu} b_{[ij]}^{\nu}
\langle  \sigma^{\nu}_{i+(jk) \rightarrow [ijk]} v \rangle
\ea
\ \ \ \ { and baryon rates}
\be
B_{[ijk]} = \sum_{\nu} b_{[ijk]}^{\nu}
\langle  \sigma^{\nu}_{i+(jk) \rightarrow [ijk]} v \rangle.
\ee

Many of these rates can be zero in a given model, specified
by a set of cross sections $\sigma^{\nu}$.
Using the above notations one can derive the following
rate equations for quarks
\be
V \frac{dN_i}{dt} \, = \, - \sum_j Q_{[ij]} N_i
\overline{N}_j
 - \sum_j Q_{(ij)} N_i N_j
 -  \sum_{jk} Q_{[ijk]} N_i N_{(jk)},
\ee

% \newpage

\noindent for diquarks
\be
V \frac{dN_{(jk)}}{dt}  =
 D_{(jk)} N_j N_k   + \sum_l D_{[jk]l} N_j \overline{N}_l
  - \sum_i D_{[ijk]} N_i N_{(jk)}
\ee
for mesons 
\be
V \frac{dN_{[ij]}}{dt}  = 
M_{[ij]} N_i \overline{N}_j  + M_{(ij)}N_iN_j 
+ \sum_k M_{[ijk]} N_i N_{(jk)}
\ee
and finally for baryons 
\be
V \frac{dN_{[ijk]}}{dt}  =
B_{[ijk]} N_i N_{(jk)} \, + \, {\rm permutations}.
\ee
Without solving this set of time-dependent differential
equations, what we eventually do, we can estimate time-averages
assuming that they are proportional to the product of
corresponding initial numbers of the reacting species.
First we integrate the rate equations for quarks, yielding
\ba
& & V  \int_0^{\tau} \,\frac{dN_i(t)}{dt} dt\,= 
 - \sum_j Q_{[ij]} \int_0^{\tau} N_i(t) \overline{N}_j(t)\ dt \nonumber \\
& - & \sum_j Q_{(ij)} \int_0^{\tau} N_i(t) N_j(t)\ dt  
 -  \sum_{jk} Q_{[ijk]}\int_0^{\tau} N_i(t) N_{(jk)}(t)\ dt.
\ea
Now we replace the time integrals of products by values scaled
by the initial numbers of (anti)quark species
\ba
&&\frac{V}{\tau}\, N_i(0) \,= 
  +  \sum_j
 Q_{[ij]} \, b_i\, N_i(0)\, \overline{b}_j\, \overline{N}_j(0)
\nonumber \\
  &+&  \sum_j  Q_{(ij)} \, b_i\, N_i(0)\, b_j \, N_j(0)
 +  \sum_{jk} Q_{[ijk]}\, b_i \, N_i(0)\, N^{{\rm eff}}_{(jk)}.
\ea
The coefficients $b_i$ are then internal variables of the model
which can be determined by solving algebraic equations.
This system of equations were closed without the diquark
contribution; their effect is taken into account by using
their effective number, $N^{{\rm eff}}_{(jk)}$, which 
represents the time integral in the original problem.

\vspace{0.2cm} 
In order to obtain $ N^{{\rm eff}}_{(jk)} $ 
we integrate the rate equation for diquarks, yielding
\ba
&& \frac{V}{\tau}\, [ N_{(jk)}(\tau) - N_{(jk)}(0) ] = 
+  D_{(jk)}  \,b_j N_j(0)\,b_k\, N_k(0) \nonumber \\
 & + & \sum_l D_{[jk]l}
\, b_j \,N_j(0) \, \overline{b}_l \, \overline{N}_l(0) 
 -  \sum_i D_{[ijk]}\, b_i \, N_i(0) N_{(jk)}(t^*)
\ea
Let us now assume, that the number of diquarks is zero both
at $ t=0 $ and at $ t= \tau $. Then from the above
equation one expresses $ N^{{\rm eff}}_{(jk)} $ as
\ba
N_{(jk)}(t^*) = \frac{
 D_{(jk)}\,b_j N_j(0)\,b_k\, N_k(0)
  +  \sum_l D_{[jk]l}
\, b_j \,N_j(0) \, \overline{b}_l \, \overline{N}_l(0) }
{ \sum_i D_{[ijk]}\, b_i \, N_i(0) }
\ea
Substituting this expression into equation(16), we can determine
the $b_i$ factors, since there are just as many independent
equation then as many $b_i $ factors we have.
Knowing the $b_i$ factors, we can integrate the differential
equations for the mesons and baryons, and thus their
number after hadronization is obtained.
\bigskip

The number of produced mesons becomes
\ba
&& N_{[ij]}(\tau) =  \frac{\tau}{V} 
               [ 
M_{[ij]} N_i(0)\overline{N}_j(0) b_i\overline{b}_j  \nonumber \\
 & + & M_{(ij)} N_i(0)N_j(0)b_ib_j 
 +  \sum_k M_{[ijk]} N_i(0) b_i N_{(jk)}^{{\rm eff}}
 ]
\ea
and the number of produced baryons:
\be
N_{[ijk]}(\tau) =  \frac{\tau}{V} 
  [   \quad B_{[ijk]} N_i(0) b_i N_{(jk)}^{{\rm eff}} 
  +  {\rm permutations} \quad  ] .
\ee
Finally we note that in order to resemble a 
coalescence model we define
\be
N_{(jk)}^{{\rm eff}} = g_B N_j(0) N_k(0) b_j b_k,
\ee
and call $g_B$ baryon suppression factor. Now every produced
hadron number is combined from initial quark number products.
\bigskip
 
In Table.1 we summarized our results on the ALCOR analysis of the
Pb+Pb data. The "experimental data" was obtained directly from the
references (see Ref. \cite{WA97S97, NA49old}) or we estimated the total
multiplicities from the published (preliminary) rapidity spectra
(see Ref. \cite{NA49S97,NA49S96}).
We assumed different values for $g_S$ to indicate differences in 
strangeness production.
Then the total $h^-$ multiplicity $(\langle h^- \rangle =680)$
was fixed by parameter $N_{q,pair}$. Finally 
we obtained  value for the baryon
suppression, $g_B$, from the estimated multiplicity of the 
${\overline Y^0}$, which is 
$\{ {\overline Y^0} \} \ \ \  \approx  8$.
Here we used the usual notation
 $Y^0 = \Sigma^0 + \Lambda^0$ and
 ${\overline Y}^0 = {\overline \Sigma}^0 + {\overline \Lambda}^0$ since these
two particles can not be distinguished experimentally \cite{NA49S97}.
\bigskip

{} From Table 1. one can obtain the following  conclusions: 
\begin{itemize}

\item The ALCOR model
can reproduce approximately the experimental data. Since these data
are preliminary, it was not required a full agreement, immediately.
\medskip

\item The best fit of the ALCOR model is in the 
region $g_S=0.22-0.24$,
which is close to the value of $g_S=0.255$ obtained in the
S+S collision. However, we did not find any further strangeness
enhancement beyond the level obtained in the S+S collision,
even it is a little bit less, as it was mentioned in Ref. \cite{NA49S97}. 
This is clearly indicated in the $Kaon/Pion$ ratio.
On the other hand the $K^+$, $K^-$ and $K^0_S$ multiplicities
can not be reproduced together, at the same time, we overpredicted 
the $K^+$.
Also we have a slight overprediction for the $Y^0$.
\medskip

\item Investigating the multi-strange hyperon ratios, one can see
that at $g_S \geq 0.24$  three of the four ratios were reproduced quite 
well by the ALCOR, on the other hand
the  ${\overline \Omega}^+$ / $\Omega^{-}$ ratio is out of the 
range of the ALCOR predictions by a factor of 2. 
Since there is a trivial 
connection among the multi-strange ratios, 
\begin{equation}
 \frac{{\overline \Omega}^+}{\Omega^-} \equiv
 \frac{{\overline \Omega}^+}{{\overline \Xi}^+} \cdot
 \frac{{\overline \Xi}^+}{\Xi^-}\cdot
 \left( { {\Omega^-} \over {\Xi^-} } \right)^{-1},
\end{equation}
we conclude that reproducing the ratios on the right hand side
with a given error do not 
automatically reproduce the   ${\overline \Omega}^+$ / $\Omega^{-}$
ratio in the left hand side. In fact we obtained a 100 \% 
disagreement  on this ratio.
\medskip

\item We nearly doubled the baryon suppression factor, $g_B$.
(It's value was $g_B^0$ for S+S collision \cite{ALCORS95}.)
This indicates that the inner microscopical
processes changed in the Pb+Pb collision, the presence
of a quark matter medium is stronger.
\end{itemize}

\newpage

\begin{table}
\begin{center}
\begin{tabular}{|c|c|c|c|c|c|c|c|} \hline
{\bf Pb+Pb} &{\bf Exp. data}& \multicolumn{6}{c|} {\bf ALCOR} \\ \hline
$g_S$                                          & ---   &
0.16 & 0.18 & 0.20 & 0.22 & 0.24 & 0.26 \\
$N_{q,pair}$                                   & ---   &
415 & 408 & 403 & 398 & 393 &  389 \\
$g_B/g_B^0$                                    & ---   &
1.6 & 1.7 & 1.8 & 1.8 & 1.9 & 1.93 \\
\hline
 $h^{-}$                                       &$ 680^{\rm \ a}  $ & 
680  & 680 & 680 & 680 & 680 & 680 \\
$K^0_{S}$                                      &$\{54 \}^{\rm \ b,c} $  &
41.5 & 46.2& 50.9& 55.1& 59.6& 63.8\\
 $K^+$                                         &$\{56\}^{\rm \ c}   $&
57.1 & 63.6& 70.3& 76.0& 82.4& 88.3\\
 $K^-$                                         &$\{32\}^{\rm \ c}   $&
25.9 & 28.7& 31.6& 34.1& 36.8& 39.4\\
\hline
 $p^+ - {\overline p}^-$                       &$\{145\}^{\rm \ a} $&
156.6&153.5&150.5&147.9&145.1&142.5\\
 $Y^0$                                         &$\{50\}^{\rm \ c}  $ & 
49.0 & 52.9& 56.7& 60.5& 63.7& 67.0\\
 ${\overline Y}^0$                             & $\ \{8\}^{\rm \ c}$ & 
 8.2 &  8.0&  7.9&  8.3&  8.1&  8.3\\
\hline
 ${\overline \Xi}^+$ / $ \Xi^-$                & $0.27 \pm 0.05^{\rm \ d}$ &  
0.39 & 0.36& 0.33& 0.33& 0.31& 0.30\\
 ${\overline \Omega}^+$ / $\Omega^{-}$         & $0.42 \pm 0.12^{\rm \ d}$ &
0.91 & 0.85& 0.81& 0.80& 0.76& 0.75\\
 $\Omega^{-}$ / $\Xi^-$                        & $0.19 \pm 0.04^{\rm \ d}$ &
0.10 & 0.11& 0.12& 0.13& 0.14& 0.15\\
 ${\overline \Omega}^+$ /  ${\overline \Xi}^+$ & $0.30 \pm 0.09^{\rm \ d}$ &
0.23 & 0.26& 0.29& 0.32& 0.35& 0.38\\
\hline
 $(K+{\overline K})/\pi$                                   & $0.13^{\rm \ c} $  & 
0.09 &0.10 & 0.11& 0.12& 0.13& 0.14\\
 $K^-/\pi^-$                                   &&
0.04 &0.04 & 0.05& 0.05& 0.06& 0.06\\
 Meson/Baryon                                  &&
4.67 &4.72 &4.77 & 4.78& 4.82& 4.84\\
\hline
\end{tabular}
\caption[]{
Hadron multiplicities are displayed for $Pb+Pb$ collision.
The values of the second column were  observed experimentally
or were estimated from the experimental data by us, 
we used the notation $\{ \ \}$.
The other columns contain the prediction of the ALCOR model 
at different strangeness production, $g_S$. }
\end{center}
\end{table}

\begin{description}
\item[] $^{\rm a}$ From \cite{NA49old}.
\item[] $^{\rm b}$ Estimated from \cite{NA49S96}.
\item[] $^{\rm c}$ Estimated from \cite{NA49S97}.
\item[] $^{\rm d}$ From \cite{WA97S97}.
\end{description}

\section*{ Acknowledgment}
Discussions with N. Bal\'azs, T. Cs\"org\H o  and
A.K. Holme  are acknowledged.
This work was supported by the National Scientific
Research Fund (Hungary), OTKA
T016206 as well as
by the U.S. - Hungarian Science and Technology
Joint Fund, No. 378/93.

\newpage

%\section*{ References}

\end{document}